\documentclass{ws-procs961x669}            % default, citations in superscript
\newcommand{\bq}{\begin{equation}}
 \newcommand{\eq}{\end{equation}}
 \newcommand{\bqn}{\begin{eqnarray}}
 \newcommand{\eqn}{\end{eqnarray}}
 \newcommand{\ba}{\begin{aligned}}
 \newcommand{\ea}{\end{aligned}}

\newcommand\be{\begin{equation}}
\newcommand\ee{\end{equation}}
\newcommand\bea{\begin{eqnarray}}
\newcommand\eea{\end{eqnarray}}
\newcommand\bseq{\begin{subequations}} %solo con amsmath
\newcommand\eseq{\end{subequations}}
\newcommand\bcas{\begin{cases}}
\newcommand\ecas{\end{cases}}

\begin{document}

\title{Solving both $H_0$ and $\sigma_8$ tensions in $f(T)$ gravity}

\author{Emmanuel N. Saridakis}

\address{National Observatory of Athens, Lofos Nymfon, 11852 Athens,
Greece\\
CAS Key Laboratory for Researches in Galaxies and Cosmology, Department of 
Astronomy, University of Science and Technology of China, Hefei, Anhui 230026, 
P.R. China\\
$^*$E-mail: msaridak@noa.gr}

\begin{abstract}
 We report how to 
alleviate both the $H_0$ and $\sigma_8$ tensions simultaneously within $f(T)$ 
gravity. In particular,   we consider the parametrization 
$f(T)=-T-2\Lambda/M_P^2+\alpha 
T^\beta$, where two out of the three parameters are independent. This model can 
efficiently fit observations solving the two tensions. To our knowledge, this is 
the first time where a modified gravity theory can alleviate both $H_0$ and 
$\sigma_8$ tensions simultaneously, hence, offering an additional argument in 
favor of gravitational modification.
\end{abstract}


\bodymatter

\section{Introduction}

It is now well established that the Universe at late times  experienced the 
transition from the matter era to the accelerated expansion phase. Although the 
simplest explanation would be the consideration of the cosmological constant, 
the corresponding problem related to the quantum-field-theoretical calculation 
of its value, as well as the possibility of a dynamical nature, led to two main 
paths of constructing extended scenarios. The first is to maintain general 
relativity as the underlying theory of gravity, and consider  new, exotic forms 
of matter that constitute the concept of dark energy  
\cite{Copeland:2006wr,Cai:2009zp}. The second is to
construct extended or modified theories of gravity, that posses general 
relativity as a low-energy limit, but which in general provide the extra 
degrees of freedom that can drive the dynamical universe acceleration  
\cite{CANTATA:2021ktz,Capozziello:2011et}.

With the accumulation of cosmological data, experimental tensions may arise 
within $\Lambda$CDM cosmology. If they were to remain under the increasing 
precision of experimental observations, they would constitute, in a statistical 
sense, clear indications of new physics beyond $\Lambda$CDM. One recently 
well-debated tension relates the value of the Hubble parameter at present
time $H_0$ measured from the cosmic microwave background (CMB) temperature and 
polarization data by the Planck Collaboration \cite{Aghanim:2018eyx} to be $H_0 
= 67.37 \pm 0.54 \ 
\mathrm{km \, s^{-1} \, Mpc^{-1}}$, to the one from local 
measurements of the Hubble Space Telescope  \cite{Riess:2019cxk} yielding $H_0 
= 74.03 \pm 1.42 \ 
\mathrm{km \, s^{-1} \, Mpc^{-1}}$. Recent analyses 
with 
the combination of gravitational lensing and time-delay effects data reported a 
significant deviation  \cite{Wong:2019kwg} at $5.3 \sigma$. Another potential 
tension concerns the measurements of the parameter $\sigma_8$, which
quantifies the gravitational clustering of matter from the amplitude of the 
linearly evolved power spectrum at the scale of $8 h^{-1} \text{Mpc}$. 
Specifically, a possible deviation was noticed between measurements of CMB and 
LSS surveys, namely, between Planck \cite{Aghanim:2018eyx} and SDSS/BOSS 
\cite{Alam:2016hwk, Ata:2017dya}.   Although these two tensions could in 
principle arise from 
unknown systematics, the
possibility of  physical origin puts the standard lore of cosmology into 
additional investigations, by pointing to various extensions beyond 
$\Lambda$CDM.

In this work we consider systematically the $H_0$ \cite{DiValentino:2020zio} and 
$\sigma_8$ \cite{DiValentino:2020vvd} tensions, 
and report how to alleviate both simultaneously within $f(T)$ gravity, based 
on the analysis of \cite{Yan:2019gbw}. We 
exploit the effective field theory (EFT) of torsional gravity, a formalism that 
allows for a systematic investigation of the background and perturbations 
separately. This approach was first developed early  
for curvature gravity \cite{ArkaniHamed:2003uy}, and recently   it was extended
to 
torsional gravity \cite{Li:2018ixg, Cai:2018rzd}.  
In order to address cosmological tensions, we identify the effects of 
gravitational modifications within the EFT on the dynamics of the background and 
of linear perturbation levels. This will allow us to construct specific models 
of $f(T)$ gravity providing adequate deviation from $\Lambda$CDM that can 
alleviate $H_0$ and $\sigma_8$ tensions.

\section{Effective field theory approach}
 
We start our analysis by the question of whether both tensions can be 
alleviated 
simultaneously via gravitational modifications in a much general framework. 
Indeed, the $H_0$ tension reveals a universe that is expanding faster at late 
times than that from a cosmological constant preferred by CMB data, while a 
lower value of $\sigma_8$ than the one of CMB most likely $\Lambda$CDM would 
imply that matter clusters either later on or less efficiently. Hence, these two 
observations seem to indicate that there might be ``less gravitational power'' 
at intermediate scales, which phenomenologically advocates a possible 
modification of gravitation. Accordingly, in this work we address the 
aforementioned question within the EFT framework for torsional gravity.

For a general curvature-based gravity, the action following the EFT approach in 
the unitary gauge, invariant by space diffeomorphisms, which expanded around a 
flat FRW metric $ds^2= -dt^2+a^2(t)\,\delta_{ij} dx^i dx^j$, is given by
\begin{align}
\label{curvEFTact}
S &= \int d^4x \Big\{ \sqrt{-g} \big[ \frac{M^2_P}{2} \Psi(t)R - \Lambda(t) - 
b(t)g^{00}
\nonumber
\\
& +M_2^4(\delta g^{00})^2 -\bar{m}^3_1 \delta g^{00} \delta K -\bar{M}^2_2 
\delta K^2 -\bar{M}^2_3
\delta K^{\nu}_{\mu} \delta K^{\mu}_{\nu} \nonumber \\
& + m^2_2 h^{\mu\nu}\partial_{\mu} g^{00}\partial_{\nu}g^{00} +\lambda_1\delta 
R^2
+\lambda_2\delta R_{\mu\nu}\delta R^{\mu\nu} +\mu^2_1 \delta g^{00} \delta R 
\big]
\nonumber \\
& +\gamma_1 C^{\mu\nu\rho\sigma} C_{\mu\nu\rho\sigma} +\gamma_2
\epsilon^{\mu\nu\rho\sigma} C_{\mu\nu}^{\quad\kappa\lambda} 
C_{\rho\sigma\kappa\lambda}
\nonumber \\
& +\sqrt{-g} \big[ \frac{M^4_3}{3}(\delta g^{00})^3 -\bar{m}^3_2(\delta 
g^{00})^2 \delta K + ... \big] \Big\} ~,
\end{align}
where $M_P = (8\pi G_N)^{-{1}/{2}}$ is the reduced Planck mass with $G_N$ the 
Newtonian constant. $R$ is the Ricci scalar corresponding to the 
Levi-Civit$\grave{\mathrm{a}}$ connection, $C^{\mu\nu\rho\sigma}$ is the Weyl 
tensor, $\delta K^{\nu}_{\mu}$ is the perturbation of the extrinsic curvature, 
and the functions $\Psi(t)$, $\Lambda(t)$, $b(t)$ are determined by the 
background evolution.

When the underlying theory also includes torsion \cite{Li:2018ixg}, one can 
generalize the EFT action as  \cite{Li:2018ixg, Cai:2018rzd}
\begin{align}
\label{actionfin}
S = & \int d^4x \sqrt{-g} \Big[ \frac{M^2_P}{2} \Psi(t)R - \Lambda(t) - b(t) 
g^{00} +
\frac{M^2_P}{2} d(t) T^0 \Big]   + S^{(2)} ~.
\end{align}
To compare with the effective action for curvature-based gravity 
\eqref{curvEFTact}, one reads that at background level there is additionally the 
zeroth part $T^0$ of the contracted torsion tensor $T^{0\mu}_{\ \mu}$, with its 
time-dependent coefficient $d(t)$.
Furthermore, the perturbation part $S^{(2)}$ contains all operators of the 
perturbation part of \eqref{curvEFTact}, plus pure torsion terms including 
$\delta T^2$, $\delta T^0\delta T^0$, and $\delta T^{\rho\mu\nu}\delta 
T_{\rho\mu\nu}$, and extra terms that mix curvature and torsion, namely, $\delta 
T\delta R$, $\delta g^{00}\delta T$, $\delta g^{00}\delta T^0$, and $\delta 
K\delta T^0$, where $T\equiv\frac{1}{4} T^{\rho \mu \nu} T_{\rho \mu \nu} + 
\frac{1}{2} T^{\rho \mu \nu
} T_{\nu \mu\rho } - T_{\rho \mu }^{\ \ \rho }T_{\ \ \ \nu}^{\nu \mu }$ is the 
torsion scalar.
Adding the matter action $S_m$ to the effective action of torsional-based 
gravity (\ref{actionfin}) and then performing variation, one obtains the 
Friedmann equations to be \cite{Li:2018ixg}:
\begin{align}
\label{f11}
H^2 &= \frac{1}{3 M_{P}^2} \big( \rho_m +\rho_{DE}^{\text{eff}} \big) ~, \\
\dot{H} &= -\frac{1}{2 M_{P}^2} \big( \rho_m +\rho_{DE}^{\text{eff}} +p_m 
+p_{DE}^{\text{eff}} \big) ~, \nonumber
\end{align}
and where
\begin{align}
\label{rhopDE}
\rho_{DE}^{\text{eff}} &= b+\Lambda -3 M_P^2 \Big[ 
H\dot{\Psi}+\frac{dH}{2}+H^2(\Psi-1) \Big] ~, \\
p_{DE}^{\text{eff}} &= b -\Lambda +M_P^2 \Big[ \ddot{\Psi} +2H\dot{\Psi} 
+\frac{\dot{d}}{2} +(H^2 +2\dot{H})(\Psi -1) \Big] ~, \nonumber
\end{align}
are, respectively, the effective DE density and pressure in the general 
torsional gravity. Moreover, we treat the matter sector as dust that satisfies 
the conservation equation $\dot{\rho}_m +3H(\rho_m +p_m) =0$, which in terms of 
redshift leads to $\rho_m = 3 M_P^2 H_0^2 \Omega_{m0}(1+z)^3$, with 
$\Omega_{m0}$ the value of $\Omega_{m}\equiv8\pi G_N\rho_m/(3H^2)$ at present.

\section{Model independent analysis}

In general, to avoid the $H_0$ tension  one needs a positive correction to the 
first Friedmann equation at late times that could yield an increase in $H_0$ 
compared to the $\Lambda$CDM scenario. As for the $\sigma_8$ tension, we recall 
that in any cosmological model, at sub-Hubble scales and through the matter 
epoch, the equation that governs the evolution of matter perturbations in the
linear regime is \cite{Anagnostopoulos:2019miu}
\begin{eqnarray}
\label{eq:delta-evolution}
\ddot{\delta}+2 H \dot{\delta} = 4 \pi G_{\mathrm{eff}} \rho_m \delta ~,
\end{eqnarray}
where $G_{\mathrm{eff}}$ is the effective gravitational  coupling given by a 
generalized Poisson equation. In general, $G_{\mathrm{eff}}$ differs from the 
Newtonian constant $G_{N}$, and thus contains information from gravitational 
modifications (note that $G_{\mathrm{eff}}=G_{N}$ in $\Lambda$CDM cosmology). 
Solving for $\delta(a)$ provides the
observable quantity $f\sigma_8(a)$, following the definitions $f(a)\equiv d\ln 
\delta(a)/d\ln a$ and $\sigma(a) = \sigma_8
\delta(1)/\delta(a=1)$. Hence, alleviation of the $\sigma_8$ tension may be 
obtained if $G_{\mathrm{eff}}$ becomes
smaller than $G_{N}$ during the growth of matter perturbations and/or if the 
``friction'' term in
\eqref{eq:delta-evolution} increases.

To grasp the physical picture, we start with a simple case: $b(t) = 0$ and  
$\Lambda(t)=\Lambda=const$ [$b$ and $\Lambda$ are highly degenerate as shown in 
\eqref{rhopDE}], while $\Psi(t)=1$. Hence, from \eqref{actionfin}, with the 
above coefficient choices, the only deviation from $\Lambda$CDM at the 
background level comes from the term $d(t)T^{0}$, and we remind that in FRW 
geometry
$T^{0}=H$ when evaluated on the background. In this case, the first Friedmann  
equation in \eqref{f11},  using for convenience the redshift $z=-1+a_0/a$ as 
the 
dimensionless variable and setting $a_0=1$, yields
\begin{eqnarray}
\label{eq:H}
H(z) = -\frac{d(z)}{4} +\sqrt{\frac{d^2(z)}{16} +H_{\Lambda \text{CDM}}^2(z)} ~,
\end{eqnarray}
where $H_{\Lambda \text{CDM}}(z) \equiv H_0 
\sqrt{\Omega_m(1+z)^3+\Omega_\Lambda}$ is  the Hubble rate in $\Lambda$CDM, 
with 
$\Omega_m=\rho_m/(3M_p^2H^2)$ the matter density parameter and primes denote 
derivatives with respect to $z$.
Accordingly, if $d<0$ and is suitably chosen, one can have $H(z\rightarrow 
z_{\rm CMB}) \approx H_{\Lambda\text{CDM}}(z\rightarrow z_{\rm CMB})$  but 
$H(z\rightarrow 0) > H_{\Lambda\text{CDM}}(z\rightarrow 0)$; i.e., the $H_0$ 
tension is solved [one should choose $|d(z)| < H(z)$, and thus, since $H(z)$ 
decreases for smaller $z$, the deviation from $\Lambda$CDM will be significant 
only at low redshift]. Additionally, since the friction term in 
\eqref{eq:delta-evolution} increases, the growth of structure gets damped, and 
therefore, the $\sigma_8$ tension is also solved [note that since we have 
imposed $\Psi=1$, then $G_{\mathrm{eff}}=G_N$ as one can verify from 
\eqref{actionfin} and \eqref{f11}; namely, the contributions from $T^0$ vanish 
at first order in perturbations].

Furthermore, for typical values that lie well within  the 
$1\sigma$ intervals of  the $H(z)$ redshift surveys, it is expected that CMB 
measurements will be sensitive to such a deviation from the $\Lambda$CDM 
scenario for nonvanishing $T^0$ at early times. Actually, the $T^0$ operator 
acts in a similar way as a conventional cosmological constant. Thus, it adds 
yet 
another new functional form to parametrize the background and leads to more 
flexibility in fitting redshift and clustering measurements.

\section{$f(T)$ gravity and cosmology}

In this section  we propose concrete models of torsional modified gravity 
that can be 
applied to alleviate the two cosmological tensions based on the torsional EFT 
dictionary. In particular, we focus on the well-known class of torsional 
gravity, namely, the $f(T)$ gravity \cite{Cai:2015emx}, which is characterized 
by the action $S=\frac{M_P^2}{2} \int d^4x e f(T)$, with $e = 
\text{det}(e_{\mu}^A) = \sqrt{-g}$ and $e^A_\mu$ the vierbein, and thus by the 
Friedmann equations 
\begin{eqnarray}
 &&H^2= \frac{\rho_m}{3M_P^2} +\frac{T}{6} -\frac{f}{6} 
+\frac{Tf_T}{3}\nonumber\\
&&\dot{H} = -\frac{1}{2 M_{P}^2} (\rho_m+p_m) 
+\dot{H}(1+f_{T}+2Tf_{TT}),
\end{eqnarray}
 with $f_{T}\equiv\partial f/\partial T$, $f_{TT} 
\equiv \partial^{2} f/\partial T^{2}$,
 where we have applied $T=6H^2$ in flat 
FRW geometry (we follow the convention of \cite{Li:2018ixg}). Therefore, $f(T)$ 
gravity can arise from the general EFT approach to torsional gravity by choosing 
$\Psi=-f_T$, $\Lambda=\frac{M_P^2}{2}(T f_T-f)$, $b=0$, $d=2\dot {f}_T$ 
\cite{Li:2018ixg}, and can restore GR by choosing $f(T)=-T-2\Lambda/M_P^2$.

The above EFT approach holds  for every $f(T)$ gravity by making a suitable 
identification of the involved time-dependent functions. For instance, we 
consider the following ansatz:
%\begin{align}
%\label{fTmodel1}
 $f(T)=-[T+6H_0^2(1-\Omega_{m0})+F(T)] $,
%\end{align}
where $F(T)$ describes the deviation from GR [note, however, that in FRW 
geometry, apart from the regular choice $F=0$, the $\Lambda$CDM scenario can 
also be obtained for the special case $F(T)=c ~ T^{1/2}$ too, with $c$ a 
constant]. Under these assumptions, the first Friedmann equation becomes
\begin{align}
\label{eq:bg3}
 T(z)+2\frac{F'(z)}{T'(z)} T(z)-F(z)= 6H^2_{\Lambda CDM}(z) ~.
\end{align}
In order to solve the $H_0$ tension, we need $T(0) = 6 H_0^2 \simeq 6 
(H_0^{CC})^{2}$, with $H_0^{CC}=74.03$ $ \mathrm{km \, s^{-1} \, Mpc^{-1}}$ 
following the local measurements \cite{Riess:2019cxk}, while in the early era of 
$z\gtrsim 1100$ we require the Universe expansion to evolve as
in $\Lambda$CDM, namely $H(z\gtrsim 1100) \simeq H_{\Lambda  CDM}(z\gtrsim 
1100)$. This requirement follows from the fact that we 
are considering modifications kicking in only at late times, and therefore, the 
results in tension from CMB analysis performed within $\Lambda CDM$ remain 
unaffected.
This implies $F(z)|_{z\gtrsim 1100}\simeq c T^{1/2}(z)$ (the value $c=0$ 
corresponds to standard GR, while for $c\neq0$ we obtain $\Lambda$CDM too). Note 
that, in this case the effective gravitational coupling is given by
\cite{Nesseris:2013jea}
\begin{eqnarray}
\label{Geff}
 G_{\mathrm{eff}}=\frac{G_{N}}{1+F_{T}} ~.
\end{eqnarray}
Therefore, the perturbation equation at linear order \eqref{eq:delta-evolution} 
becomes
\begin{equation}
\label{eq:delta-z}
 \delta'' + \left[ \frac{T'(z)}{2T(z)} -\frac{1}{1+z} \right] \delta' = \frac{9 
H_0^2 \Omega_{m0} (1+z) }{[1+F'(z)/T'(z)] T(z)} \delta ~,
\end{equation}
where $\delta\equiv\delta\rho_{m}/\rho_m$ is the matter overdensity. Since 
around the last scattering moment $z\gtrsim 1100$ the Universe should be 
matter-dominated, we impose $\delta'(z)|_{z\gtrsim 1100} \simeq 
-\frac{1}{1+z}\delta(z)$, while at late times we look for $\delta(z)$ that leads 
to an $f\sigma_8$ in agreement with redshift survey observations.

\begin{figure}[ht]
\centering
 \includegraphics[width=3.05in]{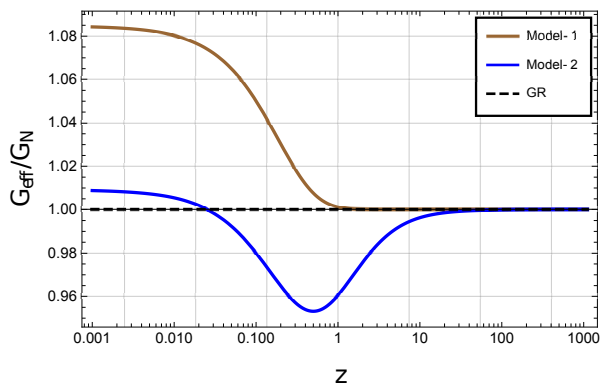}
\caption{  {\it{Redshift 
evolution of $G_{\mathrm{eff}}/G_N$ in Model-1 (brown solid line) and Model-2 
(blue solid line) and their comparison to the GR case (black dashed line).}}}
\label{fig:fTmodels}
\end{figure}

By solving \eqref{eq:bg3} and \eqref{eq:delta-z} with initial and boundary 
conditions at $z \sim 0$ and $z \sim 1100$, we can find the functional forms for 
the free functions of the $f(T)$ gravity that we consider, namely, $T(z)$ and 
$F(z)$, that can alleviate both $H_0$ and $\sigma_8$ tensions. We find two such 
forms for $F(T)$. Both models
approach the $\Lambda$CDM scenario at $z\gtrsim 1100$, with Model-1 approaching 
$F=0$ and hence restoring GR, while Model-2
approaches $F\propto T^{1/2}$, and thus it reproduces $\Lambda$CDM. In 
particular, we find that we can well fit the numerical solutions of Model-1 by
\begin{eqnarray}
\label{mod1}
 F(T) \approx 375.47 \Big( \frac{T}{6 H_0^2} \Big)^{-1.65} ~,
\end{eqnarray}
and of Model-2 by
\begin{equation}
\label{mod2}
 F(T) \approx 375.47 \Big( \frac{T}{6 H_0^2} \Big)^{-1.65} + 25 T^{1/2} ~.
\end{equation}
Note that, the first term of Model-2, which coincides with Model-1, provides a 
small deviation to $\Lambda$CDM at late times, while it decreases rapidly to 
become negligible in the early Universe. In addition, we examine 
$G_{\mathrm{eff}}$ given by \eqref{Geff} for the two models \eqref{mod1} and 
\eqref{mod2}, which are displayed in  Fig.~\ref{fig:fTmodels}. 
As expected, at high redshifts in both models, $G_{\mathrm{eff}}$ becomes $G_N$, 
recovering the $\Lambda$CDM paradigm. At very low redshifts, $G_{\mathrm{eff}}$ 
becomes slightly higher than $G_N$, increasing slightly the gravitational 
strength. This gravitational modification is in competition at late times with 
the accelerating expansion. It turns out that the effect of an increased cosmic 
acceleration with
respect to $\Lambda$CDM in our $f(T)$ gravity models dominates over the stronger 
gravitational strength in the clustering of matter. We check that both models 
can easily pass the BBN constraints (which demand \cite{Copi:2003xd}
$|G_{\mathrm{eff}}/G_N-1|\leq0.2$), as well as the ones from 
the Solar System [which demand \cite{Nesseris:2006hp} 
$|G_{\mathrm{eff}}'(z=0)/G_N|\leq10^{-3}h^{-1}$ 
and $| G_{\mathrm{eff}}''(z = 0) / G_N | \leq 10^{5} h^{-2}$].

\begin{figure}[ht]
\centering
 \includegraphics[width=3.in]{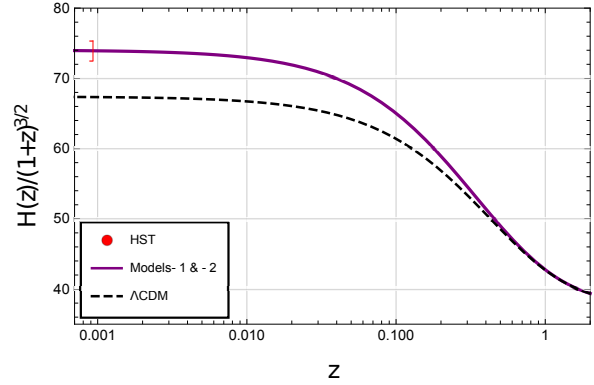}
\caption{ {\it{ Evolution of the Hubble parameter $H(z)$ in the two $f(T)$ 
models (purple solid line) and in $\Lambda$CDM cosmology (black dashed line). 
The red point represents the latest data from extragalactic Cepheid-based local 
measurement of $H_0$ provided in \cite{Riess:2019cxk}.}}}
\label{fig:H0&fs81111}
\end{figure}

\begin{figure}[ht]
\centering
\includegraphics[width=3.1in]{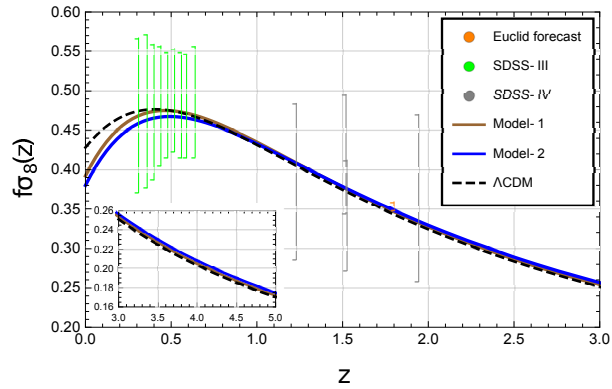}
\caption{   {\it{ Evolution 
of 
$f\sigma_8$ in Model-1 (brown solid line) and Model-2 (blue solid line)
of $f(T)$ gravity and in $\Lambda$CDM cosmology (black dashed line). The green 
data points are from BAO observations in SDSS-III DR12 \cite{Wang:2017wia}, the 
gray data points at higher redshift are from SDSS-IV DR14 
\cite{Gil-Marin:2018cgo}, while the red point around 
$\sim 1.8$ is the forecast from Euclid \cite{Taddei:2016iku}. The subgraph in 
the left bottom displays $f \sigma_8$ at high redshift $z = 3 \sim 5$, which 
shows that the curve of Model-2 is above the one of Model-1 and $\Lambda$CDM 
scenario and hence approaches $\Lambda$CDM slower than Model-1.}}}
\label{fig:H0&fs8}
\end{figure}

Now we show how Model-1 and Model-2 can alleviate the $H_0$ and $\sigma_8$ 
tension by solving the background and perturbation equations. In 
Fig.~\ref{fig:H0&fs81111} we present the evolution of $H(z)$, while in Fig. 
\ref{fig:H0&fs8} the evolution of $f\sigma_8$, for 
two $f(T)$ models, and we compare them with $\Lambda$CDM. We stress that the 
$H_0$ tension can be alleviated as $H(z)$ remains statistically consistent for 
all CMB and CC measurements at all redshifts. We remind the reader that the two 
$f(T)$ models differing merely by a term $\propto T^{1/2}$, which does not 
affect the background as explained before, are degenerate at the background 
level. However, at the perturbation level, the two models behave differently as 
their gravitational coupling $G_{\mathrm{eff}}$ differs.

We further stress that 
both models can alleviate the $\sigma_8$ tension, and fit efficiently to BAO and 
LSS measurements. Note that at high redshifts ($z\geq2$), Model-2 approaches 
$\Lambda$CDM slower than Model-1, but in a way that is statistically 
indistinguishable for present-day data. Nevertheless, future high-redshift 
surveys such as eBOSS for quasars and Euclid \cite{Laureijs:2011gra} for 
galaxies have the potential to discriminate among the predictions of $f(T)$ 
gravity and the $\Lambda$CDM scenario. Moreover, the clusters and CMB 
measurements on $\sigma_8$ are in good agreement in our models, as the CMB 
preferred values in $\Lambda$CDM get further lowered than local ones from 
rescaling $\sigma_8$ by the ratio of the growth factors in $f(T)$ gravity and 
$\Lambda$CDM. More explicitly,
\begin{equation}
\sigma_8^{f(T)}(z=0) = \frac{D^{f(T)}(z=0)}{D^{\Lambda}(z=0)} 
\frac{D^{\Lambda}(z_{\rm eff})}{D^{f(T)}(z_{\rm eff})} \sigma_8^{\Lambda}(z=0) ,
\end{equation}
where $D(z)$ is the growth factor, $f(T)$ and $\Lambda$ denote our models and 
$\Lambda$CDM respectively, and $z_{\rm eff}$ is the effective redshift of the 
measurements ($z_{\rm eff} \sim 0.1$ for clusters experiments and $z_{\rm eff} 
\sim 1100$ for CMB temperature fluctuations observations). It turns out that, as 
at high redshift, $z_{\rm eff} \sim 1100$, the growth factor is the same in 
either our $f(T)$-models or in $\Lambda CDM$, but at low redshift, $z_{\rm eff} 
\sim 0.1$, the growth factor is approximately 1.03 bigger in the later compared 
to the formers, cluster $\sigma_8$-measurements get bigger by about such an 
amount reducing the gap with CMB preferred value in this modified gravity 
scenario.

In short summary, we conclude that the class of $f(T)$ gravity: $f(T) = -T 
-2\Lambda/M_P^2 +\alpha T^\beta$, where only two out of the three parameters 
$\Lambda$, $\alpha$, and $\beta$ are independent (the third one is eliminated 
using $\Omega_{m0}$), can alleviate both $H_0$ and $\sigma_8$ tensions with 
suitable parameter choices. Moreover, such kinds of models in $f(T)$ gravity 
could also be examined through galaxy-galaxy lensing effects 
\cite{Chen:2019ftv}, strong lensing effects around black holes \cite{Li:2019lsm} 
and gravitational wave experiments \cite{Cai:2018rzd}.

\section{Conclusions}

In this work we reported how $f(T)$ gravity can alleviate 
both $H_0$ and $\sigma_8$ tensions simultaneously. Working within the EFT 
framework, torsional gravity theories can be identified as the EFT operators 
that allow us to extract the evolution equations of the background and of the 
perturbations in a model-independent manner. This allows us to address in a 
systematic way how tensions amongst the observational measurements, such as the 
ones on $H_0$ and $\sigma_8$, can be relaxed. Following these considerations, we 
constructed concrete models from specific Lagrangians in $f(T)$ gravity, 
describing cosmological 
scenarios where these tensions fade away.  Imposing initial conditions at the 
last scattering 
that reproduce the $\Lambda$CDM scenario, and imposing the late-time values 
preferred by local measurements, we reconstructed two particular forms of 
$f(T)$. These models are well described by the parametrization: $f(T) = -T -2 
\Lambda/M_P^2 +\alpha T^\beta$. To our knowledge, this is the first time where 
both $H_0$ and $\sigma_8$ tensions are simultaneously alleviated by a modified 
gravity theory.


\begin{thebibliography}{10}


  %\cite{Copeland:2006wr}
\bibitem{Copeland:2006wr}
E.~J.~Copeland, M.~Sami and S.~Tsujikawa,
{\it{Dynamics of dark energy}},
Int. J. Mod. Phys. D \textbf{15}, 1753-1936 (2006).
%doi:10.1142/S021827180600942X
%[arXiv:hep-th/0603057 [hep-th]].
%4461 citations counted in INSPIRE as of 13 Apr 2021


%\cite{Cai:2009zp}
\bibitem{Cai:2009zp}
Y.~F.~Cai, E.~N.~Saridakis, M.~R.~Setare and J.~Q.~Xia,
{\it{Quintom Cosmology: Theoretical implications and observations}},
Phys. Rept. \textbf{493}, 1-60 (2010)
%doi:10.1016/j.physrep.2010.04.001
%[arXiv:0909.2776 [hep-th]].
%641 citations counted in INSPIRE as of 30 Oct 202


 

 %\cite{CANTATA:2021ktz}
\bibitem{CANTATA:2021ktz}
E.~N.~Saridakis \textit{et al.} [CANTATA],
{\it{Modified Gravity and Cosmology: An Update by the CANTATA Network}},
[arXiv:2105.12582 [gr-qc]].
%47 citations counted in INSPIRE as of 29 Oct 2021





%\cite{Capozziello:2011et}
\bibitem{Capozziello:2011et}
S.~Capozziello and M.~De Laurentis,
{\it{Extended Theories of Gravity}},
Phys. Rept. \textbf{509}, 167-321 (2011).
%doi:10.1016/j.physrep.2011.09.003
%[arXiv:1108.6266 [gr-qc]].
%1741 citations counted in INSPIRE as of 13 Apr 2021



%\cite{Aghanim:2018eyx}
\bibitem{Aghanim:2018eyx}
  N.~Aghanim {\it et al.} [Planck Collaboration],
  {\it{Planck 2018 results. VI. Cosmological parameters}},
  Astron. Astrophys. \textbf{641}, A6 (2020)
[erratum: Astron. Astrophys. \textbf{652}, C4 (2021)].
%  arXiv:1807.06209 [astro-ph.CO].
  %%CITATION = ARXIV:1807.06209;%%

%\cite{Riess:2019cxk}
\bibitem{Riess:2019cxk}
  A.~G.~Riess, S.~Casertano, W.~Yuan, L.~M.~Macri and D.~Scolnic,
  {\it{Large Magellanic Cloud Cepheid Standards Provide a 1% Foundation for the 
Determination of the Hubble Constant and Stronger Evidence for Physics beyond 
$\Lambda$CDM}},
  Astrophys.\ J.\  {\bf 876}, no. 1, 85 (2019).
  %doi:10.3847/1538-4357/ab1422
%  [arXiv:1903.07603 [astro-ph.CO]].
  %%CITATION = doi:10.3847/1538-4357/ab1422;%%

%\cite{Wong:2019kwg}
\bibitem{Wong:2019kwg}
  K.~C.~Wong {\it et al.},
  {\it{H0LiCOW XIII. A 2.4\% measurement of $H_{0}$ from lensed quasars: 
$5.3\sigma$ tension between early and late-Universe probes}},
Mon. Not. Roy. Astron. Soc. \textbf{498}, no.1, 1420-1439 (2020).
%  arXiv:1907.04869 [astro-ph.CO].
  %%CITATION = ARXIV:1907.04869;%%
 

%\cite{Alam:2016hwk}
\bibitem{Alam:2016hwk}
  S.~Alam {\it et al.} [BOSS Collaboration],
  {\it{The clustering of galaxies in the completed SDSS-III Baryon Oscillation 
Spectroscopic Survey: cosmological analysis of the DR12 galaxy sample}},
  Mon.\ Not.\ Roy.\ Astron.\ Soc.\  {\bf 470}, no. 3, 2617 (2017).
  %doi:10.1093/mnras/stx721
%  [arXiv:1607.03155 [astro-ph.CO]].
  %%CITATION = doi:10.1093/mnras/stx721;%%

%\cite{Ata:2017dya}
\bibitem{Ata:2017dya}
  M.~Ata {\it et al.},
  {\it{The clustering of the SDSS-IV extended Baryon Oscillation Spectroscopic 
Survey DR14 quasar sample: first measurement of baryon acoustic oscillations 
between redshift 0.8 and 2.2}},
  Mon.\ Not.\ Roy.\ Astron.\ Soc.\  {\bf 473}, no. 4, 4773 (2018).
  %doi:10.1093/mnras/stx2630
%  [arXiv:1705.06373 [astro-ph.CO]].
  %%CITATION = doi:10.1093/mnras/stx2630;%%
  
   %\cite{DiValentino:2020zio}
\bibitem{DiValentino:2020zio}
E.~Di Valentino, L.~A.~Anchordoqui, O.~Akarsu, Y.~Ali-Haimoud, L.~Amendola, 
N.~Arendse, M.~Asgari, M.~Ballardini, S.~Basilakos and E.~Battistelli, 
\textit{et al.}
{\it{Snowmass2021 - Letter of interest cosmology intertwined II: The hubble 
constant tension}},
Astropart. Phys. \textbf{131}, 102605 (2021).
%doi:10.1016/j.astropartphys.2021.102605
%[arXiv:2008.11284 [astro-ph.CO]].
%120 citations counted in INSPIRE as of 31 Oct 2021

%\cite{DiValentino:2020vvd}
\bibitem{DiValentino:2020vvd}
E.~Di Valentino, L.~A.~Anchordoqui, \"O.~Akarsu, Y.~Ali-Haimoud, L.~Amendola, 
N.~Arendse, M.~Asgari, M.~Ballardini, S.~Basilakos and E.~Battistelli, 
\textit{et al.}
{\it{Cosmology intertwined III: $f\sigma_8$ and $S_8$}},
Astropart. Phys. \textbf{131}, 102604 (2021).
%doi:10.1016/j.astropartphys.2021.102604
%[arXiv:2008.11285 [astro-ph.CO]].

%\cite{Yan:2019gbw}
\bibitem{Yan:2019gbw}
S.~F.~Yan, P.~Zhang, J.~W.~Chen, X.~Z.~Zhang, Y.~F.~Cai and E.~N.~Saridakis,
{\it{Interpreting cosmological tensions from the effective field theory of 
torsional gravity}},
Phys. Rev. D \textbf{101}, no.12, 121301 (2020).
%doi:10.1103/PhysRevD.101.121301
%[arXiv:1909.06388 [astro-ph.CO]].

%\cite{ArkaniHamed:2003uy}
\bibitem{ArkaniHamed:2003uy}
  N.~Arkani-Hamed, H.~C.~Cheng, M.~A.~Luty and S.~Mukohyama,
  {\it{Ghost condensation and a consistent infrared modification of gravity}},
  JHEP {\bf 0405}, 074 (2004).
  %doi:10.1088/1126-6708/2004/05/074
%  [hep-th/0312099].
  %%CITATION = doi:10.1088/1126-6708/2004/05/074;%%

%\cite{Li:2018ixg}
\bibitem{Li:2018ixg}
  C.~Li, Y.~Cai, Y.~F.~Cai and E.~N.~Saridakis,
  {\it{The effective field theory approach of teleparallel gravity, $f(T)$ 
gravity 
and beyond}},
  JCAP {\bf 1810}, 001 (2018).
  %doi:10.1088/1475-7516/2018/10/001
%  [arXiv:1803.09818 [gr-qc]].
  %%CITATION = doi:10.1088/1475-7516/2018/10/001;%%

  
%\cite{Cai:2018rzd}
\bibitem{Cai:2018rzd}
  Y.~F.~Cai, C.~Li, E.~N.~Saridakis and L.~Xue,
  {\it{$f(T)$ gravity after GW170817 and GRB170817A}},
  Phys.\ Rev.\ D {\bf 97}, no. 10, 103513 (2018).
  %doi:10.1103/PhysRevD.97.103513
%  [arXiv:1801.05827 [gr-qc]].
  %%CITATION = doi:10.1103/PhysRevD.97.103513;%%
    


%\cite{Anagnostopoulos:2019miu}
\bibitem{Anagnostopoulos:2019miu}
  F.~K.~Anagnostopoulos, S.~Basilakos and E.~N.~Saridakis,
  {\it{Bayesian analysis of $f(T)$ gravity using $f\sigma_8$ data}},
  Phys.\ Rev.\ D {\bf 100}, no. 8, 083517 (2019).
  %doi:10.1103/PhysRevD.100.083517
%  [arXiv:1907.07533 [astro-ph.CO]].
  %%CITATION = doi:10.1103/PhysRevD.100.083517;%%
    
  
  
    %\cite{Cai:2015emx}
\bibitem{Cai:2015emx}
Y.~F.~Cai, S.~Capozziello, M.~De Laurentis and E.~N.~Saridakis,
{\it{f(T) teleparallel gravity and cosmology}},
Rept. Prog. Phys. \textbf{79}, no.10, 106901 (2016).
%doi:10.1088/0034-4885/79/10/106901
%[arXiv:1511.07586 [gr-qc]].


  
  

%\cite{Nesseris:2013jea}
\bibitem{Nesseris:2013jea}
  S.~Nesseris, S.~Basilakos, E.~N.~Saridakis and L.~Perivolaropoulos,
  {\it{Viable $f(T)$ models are practically indistinguishable from 
$\Lambda$CDM}},
  Phys.\ Rev.\ D {\bf 88}, 103010 (2013).
  %doi:10.1103/PhysRevD.88.103010
%  [arXiv:1308.6142 [astro-ph.CO]].
  %%CITATION = doi:10.1103/PhysRevD.88.103010;%%

%\cite{Copi:2003xd}
\bibitem{Copi:2003xd}
  C.~J.~Copi, A.~N.~Davis and L.~M.~Krauss,
  {\it{A New nucleosynthesis constraint on the variation of G}},
  Phys.\ Rev.\ Lett.\  {\bf 92}, 171301 (2004).
  %doi:10.1103/PhysRevLett.92.171301
%  [astro-ph/0311334].
  %%CITATION = doi:10.1103/PhysRevLett.92.171301;%%

%\cite{Nesseris:2006hp}
\bibitem{Nesseris:2006hp}
  S.~Nesseris and L.~Perivolaropoulos,
  {\it{The Limits of Extended Quintessence}},
  Phys.\ Rev.\ D {\bf 75}, 023517 (2007).
  %doi:10.1103/PhysRevD.75.023517
%  [astro-ph/0611238].
  %%CITATION = doi:10.1103/PhysRevD.75.023517;%%

%\cite{Wang:2017wia}
\bibitem{Wang:2017wia}
  Y.~Wang, G.~B.~Zhao, C.~H.~Chuang, M.~Pellejero-Ibanez, C.~Zhao, 
F.~S.~Kitaura 
and S.~Rodriguez-Torres,
  {\it{The clustering of galaxies in the completed SDSS-III Baryon Oscillation 
Spectroscopic Survey: a tomographic analysis of structure growth and expansion 
rate from anisotropic galaxy clustering}},
  Mon.\ Not.\ Roy.\ Astron.\ Soc.\  {\bf 481}, no. 3, 3160 (2018).
  %doi:10.1093/mnras/sty2449
%  [arXiv:1709.05173 [astro-ph.CO]].
  %%CITATION = doi:10.1093/mnras/sty2449;%%

%\cite{Gil-Marin:2018cgo}
\bibitem{Gil-Marin:2018cgo}
  H.~Gil-Mar\'{i}n {\it et al.},
  {\it{The clustering of the SDSS-IV extended Baryon Oscillation Spectroscopic 
Survey DR14 quasar sample: structure growth rate measurement from the 
anisotropic quasar power spectrum in the redshift range $0.8 < z < 2.2$}},
  Mon.\ Not.\ Roy.\ Astron.\ Soc.\  {\bf 477}, no. 2, 1604 (2018).
  %doi:10.1093/mnras/sty453
%  [arXiv:1801.02689 [astro-ph.CO]].
  %%CITATION = doi:10.1093/mnras/sty453;%%

 

%\cite{Taddei:2016iku}
\bibitem{Taddei:2016iku}
  L.~Taddei, M.~Martinelli and L.~Amendola,
  {\it{Model-independent constraints on modified gravity from current data and 
from the Euclid and SKA future surveys}},
  JCAP {\bf 1612}, 032 (2016).
  %doi:10.1088/1475-7516/2016/12/032
 % [arXiv:1604.01059 [astro-ph.CO]].
  %%CITATION = doi:10.1088/1475-7516/2016/12/032;%%

%\cite{Laureijs:2011gra}
\bibitem{Laureijs:2011gra}
  R.~Laureijs {\it et al.} [EUCLID Collaboration],
  {\it{Euclid Definition Study Report}},
  arXiv:1110.3193 [astro-ph.CO].
  %%CITATION = ARXIV:1110.3193;%%
  
 

%\cite{Chen:2019ftv}
\bibitem{Chen:2019ftv}
  Z.~Chen, W.~Luo, Y.~F.~Cai and E.~N.~Saridakis,
  {\it{New test on General Relativity using galaxy-galaxy lensing with 
astronomical surveys}},
Phys. Rev. D \textbf{102}, no.10, 104044 (2020).
%  arXiv:1907.12225 [astro-ph.CO].
  %%CITATION = ARXIV:1907.12225;%%

%\cite{Li:2019lsm}
\bibitem{Li:2019lsm}
  S.~Yan, C.~Li, L.~Xue, X.~Ren, Y.~F.~Cai, D.~A.~Easson, Y.~Yuan and H.~Zhao,
  {\it{Testing the equivalence principle via the shadow of black holes}},
  Phys. Rev. Res. \textbf{2}, no.2, 023164 (2020).
%  [arXiv:1912.12629 [astro-ph.CO]].

 
 
  \end{thebibliography}
\end{document}